\documentclass[12pt,a4paper]{article}
\usepackage{epsfig,amsfonts}
\newcommand \blackboardrrm{\mathchoice
{\rm I\kern-0.21 em{R}}{\rm I\kern-0.21 em{R}}
{\rm I\kern-0.19 em{R}}{\rm I\kern-0.19 em{R}}}

\newcommand \blackboardzrm{\mathchoice
{\rm Z\kern-0.32 em{Z}}{\rm Z\kern-0.32 em{Z}}
{\rm Z\kern-0.28 em{Z}}{\rm Z\kern-0.28 em{Z}}}

\newcommand \be  {\begin{equation}}
\newcommand \ee  {\end{equation}}
\newcommand \lan {\langle}
\newcommand \ran {\rangle}

\renewcommand \d {{\mbox d}}

\newcommand \dof {\mbox{DF}}

\begin{document}

\title{Critical properties of Ising model
on Sierpinski fractals. A finite size scaling analysis approach.}

\author{Jos\'e M. Carmona$^a$, Umberto Marini 
Bettolo Marconi$^b$,\\
Juan J. Ruiz-Lorenzo$^c$    and Alfonso Taranc\'on$^a$\\[0.5em]
$^a$ {\small Departamento de F\'{\i}sica Te\'orica,}
{\small Universidad de Zaragoza,}\\
{\small   Plaza S. Francisco s/n, 50009 Zaragoza (Spain)}\\[0.3em]
$^b$ {\small Dipartimento di Matematica e Fisica and Istituto Nazionale}\\ 
{\small di Fisica della Materia, 
Universit\`a di Camerino, }\\
{\small   \ \ Via Madonna delle Carceri, 62032, Camerino (Italy) }\\[0.3em]
$^c$ {\small Dipartimento di Fisica and Istituto Nazionale}\\
     {\small  di Fisica Nucleare (Sez. Roma-I), 
Universit\`a di Roma}
   {\small {\em La Sapienza},}\\
{\small   \ \  P. A. Moro 2, 00185 Roma (Italy)}\\[0.3em]
{\small \tt carmona,tarancon@sol.unizar.es}\\
{\small \tt umberto.marini.bettolo@roma1.infn.it}\\
{\small \tt ruiz@chimera.roma1.infn.it}\\[0.5em]}

\date{\today}

\maketitle

\begin{abstract}

 The present paper focuses on the order-disorder transition of an
 Ising model on a self-similar lattice.  We present a detailed
 numerical study,
 based on the Monte Carlo method in conjunction with the finite size
 scaling method, of the critical properties of the Ising model on 
 some two dimensional deterministic fractal lattices
 with different Hausdorff dimensions. Those with finite ramification
 order do not display ordered phases at any finite temperature, whereas
 the lattices with infinite connectivity show genuine critical
 behavior.  In particular we considered two Sierpinski carpets
 constructed using different generators and characterized by
 Hausdorff dimensions
 $d_H=\ln 8/\ln 3 = 1.8927..$ and $d_H=\ln 12/\ln 4 = 1.7924..$, respectively.
  The data show in a clear way the existence
 of an order-disorder transition at finite temperature in both 
Sierpinski carpets.
 By performing several Monte Carlo simulations at different temperatures 
and on lattices of increasing size in conjunction with a finite size
scaling analysis, we were able to determine numerically the critical
exponents in each case and to provide an estimate of their errors. 
 Finally we considered the hyperscaling relation and found indications that 
it holds, if one assumes that the relevant dimension in this case
is the Hausdorff dimension of the lattice.

\end{abstract}



%

\thispagestyle{empty}

\newpage

\section{Introduction}

The present understanding of phase transitions has greatly benefited
from the study of the spin-lattice models, perhaps the simplest examples of
extended systems showing non trivial cooperative behavior, such as 
spontaneously symmetry breaking. In most cases one is interested in studying
systems whose geometrical properties are regular so that one
assumes that the spins occupy the cells of a regular Bravais lattices. 
Near the critical point, i.e. when the correlation length
is much larger than the lattice spacing, the influence of 
the lattice structure becomes negligible 
and only the embedding dimension, the number of components of the order 
parameter together with its symmetry, and the nature of the couplings
concur to determine the values of the critical exponents.

Such universal behavior is absent if the lattice is a fractal, because
the translational invariance is replaced by the much weaker dilation
invariance.  Notwithstanding the intense activity on various physical
problems in a space which instead of being Euclidean is a fractal
lattice, the issue of the phase transitions and of the critical
properties on self-similar supports has been rarely addressed
\cite{note,spheric,Io}. Unlike critical phenomena in spaces of integer
dimension those occurring in self-similar geometries have not been
explored so far, apart from some isolated cases.  One expects that
the lack of translational invariance plays a crucial role in phase
transitions on fractal supports.  These systems besides serving in
practice to model natural materials such as porous rocks, aerogels,
sponges etc., provide a geometric realization of non integer Hausdorff
dimension.  Therefore they offer a possibility of testing the
$\epsilon$-expansion technique for $\epsilon$ not integer. Finally,
one can explore systems whose geometrical dimension 
is very near to its
lower critical dimension (in the Ising Model is one: i.e. one is the
larger (integer) dimension in which the system does not 
have a phase transition at finite temperature).

In the present study we shall consider lattices obtained by removing
sites from a square lattice. If the diluted lattice is generated by a
sequence of random deletions one obtains the so-called site diluted
Ising model (SDIM), whose phase diagram has been studied (see for
instance \cite{PARU,DILU4,DILU2}).  In this case, one finds that the
critical temperature $T_c$ tends to zero as the probability $p$ of 
having a site tends to the percolation threshold $p_c=0.592746$ (on a
square lattice). Notice that only at the percolation
threshold $p_c$ the lattice manifests true self-similarity (and fractal
dimension 1.8958~\cite{STAUFFER}), but the
phase transition occurs only at $T=0$.  Thus to observe simultaneously
genuine criticality and self-similarity we consider a non-stochastic
lattice, the so called Sierpinski carpet (SC).  As pointed out some
years ago, one can decide whether a lattice is able to support
an order-disorder transition of the Ising type by looking at its order
of ramification ${\cal {R}}$, which is finite if after eliminating a
finite number of bonds one can isolate an arbitrarily large
sublattice. Only if ${\cal {R}}=\infty$ a phase transition
occurs. Thus the SC displays a critical point at finite temperature,
while the Sierpinski gasket does not, 
being characterized by a finite
${\cal {R}}$. This is a manifestation of the fact that, when
considering non regular structures, the embedding dimension alone is
not sufficient to determine the phase behavior.

The deterministic SC we consider, 
in spite of having a lower Hausdorff dimension than that of the SDIM, is always
above the percolation threshold by construction. One can thus
observe the interplay between fractal geometry and thermodynamics.

 In the present work we shall investigate the critical behavior of the
Ising model on two classes of fractal lattices, the two dimensional
Sierpinski gasket and the two dimensional Sierpinski carpet. The first
lattice, with Hausdorff dimension $d_H=\ln 15/\ln 5 =
1.682606...$, can be studied analytically and an exact real space
renormalization group treatment rules out the possibility of a finite
ordering temperature.  The Sierpinski carpet 
instead displays a genuine transition at
finite temperature that we have characterized numerically
for two different fractal dimensions, $d_H=\ln 8/\ln 3$, and
$d_H=\ln 12/\ln 4$.

 In statistical mechanics it is usually assumed that in the infinite
volume limit ($V\to \infty$) the average of a given observable
calculated over a subvolume sufficiently large compared with the bulk
correlation length yields a result independent of the particular
choice and location of the cell within the sample and moreover that
this average converges to the average value over the whole sample.  In
the         fractal lattices that
we consider such property breaks   down below the critical
point because the system cannot be regarded as a uniform material over
length scales larger than the correlation length, due to the presence
of voids of all sizes which render the system effectively not uniform.
Even the correlation length depends on the position. However we shall
show that this system has a second order phase transition 
and that means the existence of a
divergent length scale $\xi_\infty$ in the thermodynamic
limit. Although the fractal is not homogeneous, we postulate the
existence of this ``average'' correlation length at least near
criticality. To assume the existence of such a length is in some sense
similar to what happens when studying the diffusion problem on a
lattice. There, of course, a walker does not diffuse with the same law
from any point of the lattice, but one can still define a diffusive
type of behavior of the type $R\sim t^{1/d_w}$ by averaging over all
possible initial positions of the walker.

We outline the plan of the paper:
in section 2 we introduce the lattices and define the observables of the Ising
model which will be measured in the simulations, 
and give some details about the numerical procedure
employed to obtain the statistical averages. In section 3 we illustrate
some analytical results 
concerning lattices with finite ramification order,
which were 
obtained by means of an exact Migdal-Kadanoff decimation procedure.
In section 4 we recall briefly the statement of
the finite size scaling method.
In section 5 we illustrate the results
of the simulations for the fractal characterized by
$d_H=\ln 8/\ln 3$ 
and discuss at some length the details of the
data analysis on the basis of which we determined the values
of the various critical exponents and of the critical coupling.
In section 6 we illustrate
the results of the simulations for the
fractal with $d_H=\ln 12/\ln 4$, 
with special emphasis to the hyperscaling relation.
Finally in 7 we present the conclusions.

\section{The Model and Observables}

\begin{figure}[tb]
\centerline{\epsfig{figure=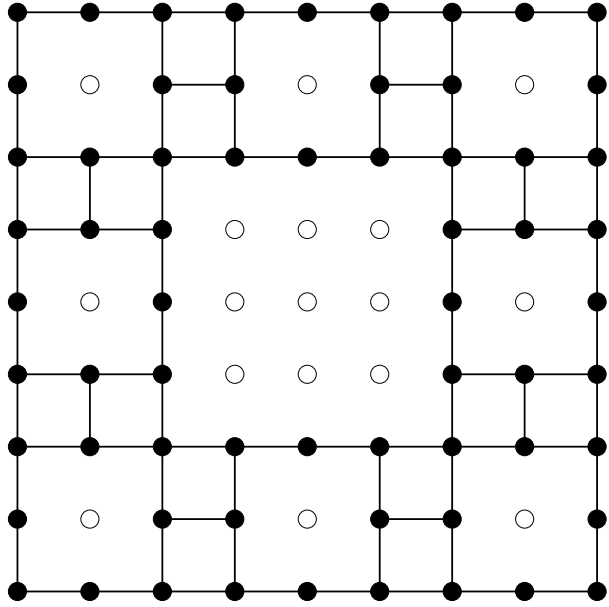,angle=90,width=80mm}}
\caption{SC of dimension $d_H=\ln 8/\ln 3$ for $L=9$. 
Filled circles represent cells to which Ising spins are assigned,
whereas empty circles
represent cells which have been eliminated from
the original square lattice. The links represent the interactions 
among spins.}
\label{FIG:FRACTAL}
\end{figure}

Let us introduce  
the lattice models that we shall consider in numerical studies.
We have first considered a SC, named hereafter fractal A, 
constructed starting 
from a square lattice of $L \times L$ cells with $L=3^n$,
dividing it into $3 \times 3$ blocks of equal size and discarding the 
cells contained
in the central block. Divide again each of the remaining blocks into 
$3 \times 3$ sub-blocks and discard all the central elements. Carry on this 
procedure until the smallest sub-block contains a single cell. 
The resulting structure 
is formed by $V$ cells, where $V$ is related to the linear 
dimension through the Hausdorff dimension $d_H$ as $V=L^{d_H}$.
In this case, $d_H=\ln 8/\ln 3=1.892789...$
Of course one can build many different SC with the same fractal dimension,
but having different distributions of voids. This can be quantified
by means of the so called lacunarity \cite{voids}. In the present work
we stick to the symmetrical fractal shown in Fig. \ref{FIG:FRACTAL}.
We define bonds between the remaining cells. The number of bonds $N$ is
\begin{equation}
N={8\over5}V+{2\over5}L,
\end{equation}
so that $N$ is proportional to $V$ in the thermodynamic limit. Note that
this fractal has ${\cal {R}}=\infty$.

 For the fractal B,  
the construction follows the same procedure, but now one considers 
$4 \times 4$ blocks and discards the four central ones. This lattice has
fractal dimension $d_H=\ln 12/\ln3 = 1.792481...$ and in this case the
relation between the number of bonds $N$, the linear dimension $L$ and
the number of cells $V$ is
\begin{equation}
N={3\over2}V+{1\over2}L.
\end{equation}

 Now, at each unit cell we assign a dichotomic spin variable $\sigma=\pm 1$.
The total energy of the system with ferromagnetic interactions is defined as
\begin{equation}
{\cal H}= -\sum_{<i,j>} \sigma_i \sigma_j \ ,
\label{eq:eq1}
\end{equation}
where $<i,j>$ are nearest neighbor cells. 
The energy normalized to the number of links is a quantity that has a
well-defined thermodynamic limit, and the specific heat is:
\begin{equation}
C=N \left(\lan E^2\ran-\lan E\ran^2\right) \ .
\end{equation}

Since our primary target is to obtain informations about the
critical behavior of the model
we first locate the critical point and then we extract the critical exponents.
To achieve this goal we introduce 
the observables which are relevant. These are defined as  follows:
the intensive magnetization $M$
\begin{equation}
M=\frac{1}{V} \sum_i \sigma_i \ ,
\end{equation}
the isothermal susceptibility:
\begin{equation}
\chi=V \left(\lan M^2\ran-\lan|M|\ran^2\right) \ ,
\label{susc}
\end{equation}
and  the Binder cumulant
\begin{equation}
U=\frac{1}{2}\left(3-\frac{\lan M^4\ran}{\lan M^2\ran^2}\right) .
\label{cumulant}
\end{equation}

\section{Migdal-Kadanoff Method}

The Migdal-Kadanoff (MK) decimation 
is an approximation very suitable for
analyzing low dimensional systems. MK is exact in one dimension, but
lacks predictability in higher
dimensions. The starting point is the
formula for the variation of the inverse temperature ($\beta$) with
the scale factor in the MK approximation in a system of dimensionality
$D$~\cite{PaMa,formula},
\begin{eqnarray}
\frac{\d \beta}{\d t}&=& g(\beta) \ , \\ 
g(\beta) &=& (D-1) \beta + 
\sinh(\beta) \cosh(\beta) \ln\left[\tanh( \beta) \right] \ ,
\end{eqnarray}
where $t$ is the logarithm of the scale factor in the block
construction.

The zeroes of $g(\beta)$ yield the inverse critical temperatures of the
system. Also, it is possible to compute the $\nu$ exponent from
\be
\frac{1}{\nu} = \left. \frac{\d g(\beta)}{\d \beta}
\right|_{\beta=\beta_c} \ .
\ee

Such a formula is very interesting because one can obtain the behavior of
the critical exponents and the critical temperature near the lower
critical dimension of the Ising model; for instance when $D \to 1$
\be
\beta_c \simeq\frac{1}{2(D-1)} \ , \,\,\, \nu \simeq \frac{1}{D-1} \ .
\ee

If one assumes that the dimensionality of the system 
corresponds to the Hausdorff dimension (i.e. $D=d_H=1.892789...$)
one obtains $\beta_c\simeq 0.56$ and $\nu \simeq 1.12$. 
Numerically (solving $g(\beta_c)=0$ and
$1/\nu=g^\prime(\beta_c)$ for $D=1.89$, that corresponds to fractal A) 
one obtains:
\be
\beta_c=0.5120 \ , \,\,\, \nu=1.409 \ .
\ee
For the fractal B ($D=1.79$) the results are
\be
\beta_c=0.5906 \ , \,\,\, \nu=1.511 \ .
\ee

Obviously this approach does not take into account the ramification
and lacunarity of the model. Nonetheless in this section we have
obtained a first guess of the critical temperature and of the $\nu$
exponent for a system with the same fractal dimension that the
Sierpinski carpet (in the most symmetric version) that we will study
numerically in the next sections.

\subsection{Exact solution of the Sierpinski gasket}

The decimation can be carried out exactly on the Sierpinski gasket,
recovering the same lattice after a
decimation transformation.

We express the Boltzmann weight in the usual way
\be
\exp(\beta s_1 s_2) = \cosh(\beta) \left[1 +s_1
s_2 \tanh(\beta)  \right] \ ,
\ee
where we have denoted by $s_1$ and $s_2$ two generic spins 
belonging to the lattice.
We denote by $\mu$ the spins that we decimate and by $\sigma$ the rest
of the spins.
\begin{figure}[tb]
\centerline{\epsfig{figure=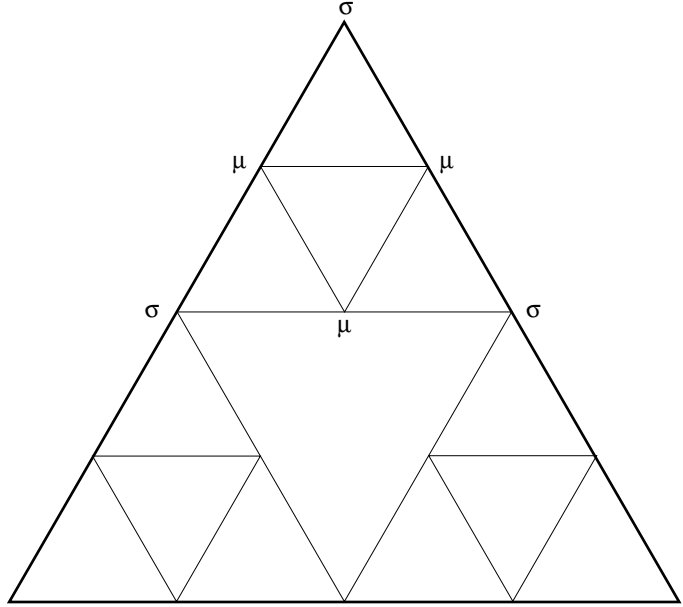,angle=270,width=90mm}}
\caption{Sierpinski gasket (SG). The spins live in the sites of the
construction. We have marked with $\mu$ the spins that are decimated. See
the text for more details.}
\label{FIG:SG}
\end{figure}

The decimation over the $\mu$ spins is the following (see Fig. \ref{FIG:SG})
\begin{eqnarray}\nonumber
&&\sum_{\mu_1,\mu_2,\mu_3=\pm 1}
\exp\left[\beta (\sigma_1 \mu_1 +
\sigma_1 \mu_2 + \mu_1 \mu_2 + \mu_1 \sigma_2 + \mu_1 \mu_3 + \mu_2
\mu_3 + \mu_2 \sigma_3 + \mu_3 \sigma_3 + \mu_3 \sigma_2)\right]  \\ \nonumber
&=& F \sum 
( 1 + k \sigma_1 \mu_1)
( 1 + k \sigma_1 \mu_2) 
(1+k \mu_1 \mu_2) 
(1 +k \mu_1 \sigma_2) 
(1 + k \mu_1 \mu_3) \\ \nonumber
&\times&(1+k \mu_2 \mu_3)
(1+k \mu_2 \sigma_3) 
(1+k \mu_3 \sigma_3) 
(1+k \mu_3 \sigma_2)\\ \nonumber
&=&8 (1+k)^3 (1+k^2)(1-3 k + 5 k^2 -3 k^3 + k^4) \\
&\times&
 \left[1
+\frac{k^2}{ 1-3 k + 5 k^2 -3 k^3 + k^4}( \sigma_1 \sigma_2+ \sigma_1
\sigma_3 + \sigma_2 \sigma_3) \right]
\label{MK_1}
\end{eqnarray}
where we have defined $k\equiv \tanh(\beta)$ and $F=\cosh^9(\beta)$.

The Sierpinski gasket is self-similar under this decimation
transformation and we can see that eq. (\ref{MK_1}) corresponds to a
nearest neighbor Ising interaction with a
renormalized coupling ($\beta_R$): i.e.
\begin{eqnarray}\nonumber
\exp\left[\beta_R (\sigma_1 \sigma_2 + \sigma_2 \sigma_3 + \sigma_3
\sigma_1)   \right]
&=& F^\prime (1 + k_R \sigma_1 \sigma_2) (1 + k_R \sigma_1 \sigma_3)
(1 + k_R \sigma_2 \sigma_3) \\
&=& F^\prime (1+k_R^3)\left[1 + \frac{ k_R ( 1 +k_R)}{1 + k_R^3} 
(\sigma_1 \sigma_2+ \sigma_1
\sigma_3 + \sigma_2 \sigma_3) \right] \ .
\label{MK_2}
\end{eqnarray}
and we have
denoted with $k_R=\tanh(\beta_R)$ the renormalized $k$ and 
$F^\prime=\cosh^3(\beta_R)$.  
We finally arrive at the following exact recursion (by matching the
coefficient of $(\sigma_1 \sigma_2 + \sigma_2 \sigma_3 + \sigma_3
\sigma_1)$ of eqs. (\ref{MK_1}) and (\ref{MK_2}))
\be
\frac{ k^2}{1-3 k + 5 k^2 -3 k^3 + k^4} = \frac{ k_R ( 1 +
k_R)}{1 + k_R^3} \ .
\ee
The critical points correspond to 
the solutions of the previous recursion
formula. The solutions are $k=0$ and $k=1$, i.e. are 
$\beta=0$ and $\beta=\infty$: i.e. there is no phase transition at
any finite temperature. From Fig. \ref{FIG:SG} it is clear that
$\cal R$ remains finite even when the linear size diverges.

\section{Finite Size Scaling Method (FSS)}

For a finite system whose typical linear size is 
$L$, which is assumed to be much larger than the
lattice spacing $a$, the
finite size scaling hypothesis~\cite{Brezin,Barber}
postulates that 
upon approaching the critical point the average value of a given observable,
$P$, depends on the size and on the temperature through the
following scaling relation:

\begin{equation}
{P_L(t)\over P_\infty(t)}=f\left(L\over\xi_\infty(t)\right),
\end{equation}
where $T_c$ and $t$ are respectively the critical and the reduced
temperature $t=|T-T_c|/T_c$ of the system,  
and $P_\infty$ represents the value of $P$ in the infinite volume
limit.
If $P_\infty(t)$ behaves as $t^{-\rho}$ when $t\to 0$, since 
the correlation length diverges as $\xi_{\infty}\sim t^{-\nu}$,
it is clear that $P_L(t)$ will saturate when the $\xi_{\infty}$
becomes comparable with $L$.
This can be formalized by writing:
\begin{equation}
P_L(t)=L^{\rho/\nu}g\left(tL^{1/\nu}\right),
\end{equation}
where the scaling function $g(x)$ 
has the limiting behavior $g(x)={\rm const}$ as 
$x\to 0$ and $g\sim x^{-\rho}$ as $x\to \infty$.

Thus at the critical point one can write the following finite
size scaling formulae for the observables (see for
instance \cite{DILU4}):
\be
\chi(L,t=0) \propto L^{\gamma/\nu} \left[ 1 +O(L^{-\omega}) \right],
\ee
\be
C(L,t=0) \propto L^{\alpha/\nu} \left[ 1 +O(L^{-\omega}) \right],
\ee
where $\omega$ is the correction-to-scaling exponent (it is just the
slope of the field theoretical $\beta$-function at the fixed point or
equivalently 
the largest irrelevant scaling exponent in the Wilson renormalization group),
and $t$ is the reduced temperature of the model.

For  non dimensional quantities, denoted by $A$, 
in the proximity of the critical point ($t \ll 1$) one has the scalings:
\be
A(L,t) = f_A(L^{1/\nu} t) +L^{-\omega} g_A(L^{1/\nu} t) + O(L^{-2 \omega}).
\ee
where $f_A$ and $g_A$ are scaling functions. Their derivatives 
with respect to $\beta$ behave as:
\be
\frac{\d A}{\d \beta}(L,t=0) \propto L^{1/\nu} \left[ 1 +O(L^{-\omega})
\right].
\ee
In particular in the present work we have chosen $A$ 
to be the Binder cumulant $U$ and $A=\ln
\lan |M| \ran$.

The $\omega$ exponent can be calculated using the effective critical
exponents that are extracted from the peaks of the observables at lattice
sizes $L$ and $sL$. Let us take the susceptibility, for example. We
obtain the effective $\frac{\gamma}{\nu}(L,sL)$ as
\be
\frac{\gamma}{\nu}(L,sL)=\frac{1}{\ln s} \ln\frac{\chi(s L)}{\chi(L)}\ .
\label{omega}
\ee 
Using the scaling formulae and working in the large lattices, 
we obtain
\be
\frac{\gamma}{\nu}(L,sL)=\frac{\gamma}{\nu} + BL^{-\omega},
\label{fitomega}
\ee
where $\gamma/\nu$ is the infinite volume extrapolation of the
ratio $\gamma/\nu(L,sL)$ measured on finite lattices. We obtain $\omega$
and $\gamma/\nu$ by fitting the data to formula (\ref{fitomega}).

In order to determine the critical 
temperature, let us call $\beta^O_c(L)$
the value of $\beta$ where the observable $O$ (with positive
dimension) displays a maximum. In order to assess its location
several simulations at different
values of $\beta$ are needed in principle, and even doing so
its precise determination is hard.
However, the spectral density method (SDM)~\cite{SDM} renders such task easier
and much faster.
In fact, by using the data from a single simulation 
at a given temperature, say $\beta_0$, one can obtain information
about some $\beta_1$
in the neighbourhood of $\beta_0$. With SDM one usually
gains one order of magnitude in the accuracy of the location of the peak.

The idea is to write  the partition function under the form:
\begin{equation}
Z\propto \sum_{\mathrm{\{confs\}}} e^{-\beta {\cal H}(\sigma)}=
\sum_E\sum_{\mathrm{\{confs\}}} e^{-\beta {\cal H}(\sigma)}\, 
\delta({\cal H}-E)=
\sum_E e^{-\beta E}N(E),
\end{equation}
where ${\cal H}(\sigma)$ is the Hamiltonian as a function of the configuration
$\{\sigma\}$ and $N(E)$ is the energy density of states, which can be 
obtained from the number of times that the value 
$E$ of the energy is generated
during a Monte Carlo simulation.
Given the  density of states at
$\beta=\beta_0$,  the value of a certain operator $O$ at 
$\beta_1\neq\beta_0$ is given by
\begin{equation}
\langle O(\beta_1)\rangle = 
\frac{\sum_E O(\beta_0,E) \,N(E)\,e^{-(\beta_1-\beta_0)E}}
{\sum_E N(E) e^{-(\beta_1-\beta_0)E} }.
\end{equation}
This extrapolation is exact. The problem is that for large 
deviations
$\Delta\beta=\beta_1-\beta_0$ the errors in the extrapolation are very
large, and then one has to restrict oneself to small $\Delta\beta$ values
(of order $1/\sqrt{V}$).

In $d\ge2$ the value $\beta_c^O(L)$ hardly changes with $O$,
and with the help of the SDM one can reach
the peak of every observable
with a single simulation at a certain $\beta$. We found that this is
not the case 
on fractal lattices with
$d_H<2$ because the transition region turns out to be very
large, and sometimes 
we were forced to perform different simulations at
different points, depending on the observables, in order to get their
corresponding peaks.

\begin{figure}[htb]
\centerline{\epsfig{figure=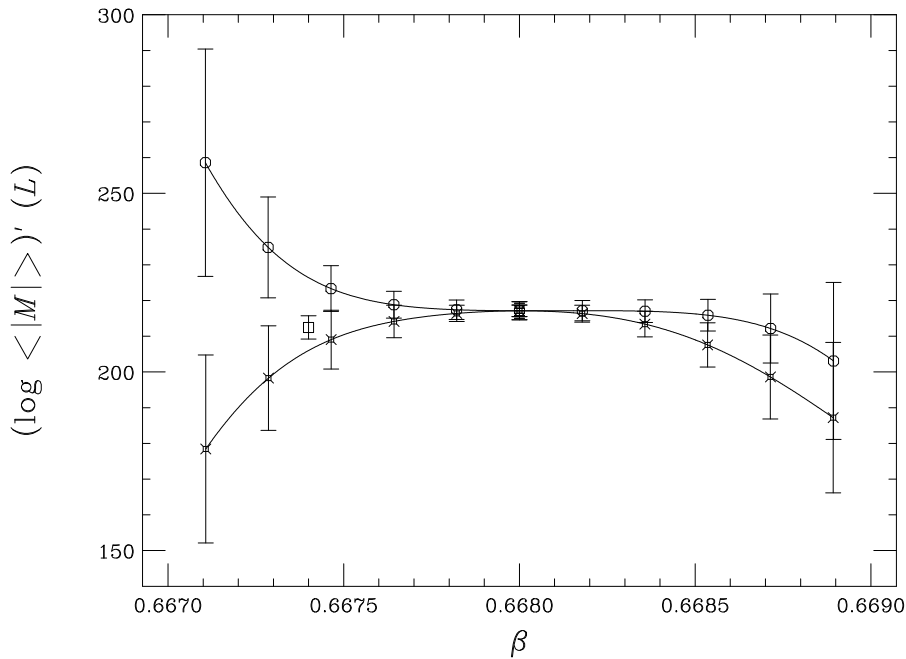,angle=90,width=120mm}}
\caption{SDM extrapolation for the derivative of $\ln \lan |M|\ran$
 in a $L=2187$ lattice.
Circles are results relative to Wolff algorithm, 
crosses are relative to SW. The extra point refers 
to another simulation (not a SDM extrapolation). The region where the
SDM extrapolation has small errors is very small.}
\label{FIG:DLOGM2187}
\end{figure}

To compute thermal averages we have employed a 
combination of $m$ steps of the classical Metropolis algorithm followed by
$n$ steps of the Wolff single-cluster algorithm
\cite{WOLFF}, which
gives a very short autocorrelation time.
We checked that
such method is sufficient to ensure ergodicity. 
However, the Wolff algorithm is unable to flip clusters of
intermediate size, which might be important in our model, where there are
domains of spins at every scale. That is why we have applied in 
addition a Swendsen-Wang algorithm \cite{SW}, 
which is more consuming  in computing time
than the Wolff algorithm, but
is able to flip domains of spins of every scale. In fact, we have
compared the results of the two different procedures. 
Fig. \ref{FIG:DLOGM2187}
shows that the two methods give compatible results in a region
around the simulation point, where the SDM extrapolation is valid
(away from this region, the SDM extrapolation gives large errors and
it is therefore not reliable, so that it is not surprising that it
gives different results for the two algorithms).
We also note the
narrow range of validity of the SDM which is used to extrapolate
the numerical data,
which makes even more troublesome the problem of finding the peaks of
observables in order to perform FSS.  As far as the
observables directly measured are concerned one can hardly see
differences in the two simulation algorithms; however, discrepancies
exist when one looks at the derivatives of the measured quantities.

\section{Numerical Results for the fractal A}

We  focus now attention on the SC of fractal dimension $d_H=\ln 8/\ln 3$
described in section 2 (fractal A), and provide
evidence of the existence of a phase 
transition at a finite value of $\beta$. 
The behavior of the SC should interpolate between the $d=1$ and $d=2$
cases. In $d=1$, $\beta_c$ diverges. In fact, with a similar analysis 
of the partition function to the one used in the Appendix, it can be
analytically calculated that $\xi(\beta)\sim|\ln\tanh(\beta)|^{-1}$,
which means that, when $\beta\to\infty$, $\xi(\beta)\sim e^{2\beta}$.
If we define $\beta_c(L)$ as the value of $\beta$ for which $\xi\simeq L$,
then $\beta_c(L)\sim\ln L$ in $d=1$. 
This means a smooth growth; it is therefore 
necessary to pay attention in order to 
demonstrate that $\beta_c(L)$ does not diverge with $L$ in the case
under scrutiny.

We performed different simulations for lattice sizes with
$L=27,81,243,729$ and $2187$, and concluded that there are several
evidences of the occurrence of a phase transition in the Ising model
on this
Sierpinski carpet.  Not only the average magnetization changes
from $0$ to $1$ going from the disordered to the ordered phase, but we
also observed that some observables, shown below, present peaks
increasing as powers of the size, $L$.  The most stringent criterion
to locate the critical coupling, $\beta_c$ is to study the crossing of
the Binder cumulants relative to different volumes. 
In $d=1$, where there is not a phase transition at a finite value
of $\beta$, the cumulants cross each other at $\beta=\infty$. This is not 
the case in this model. The positions of
the intersections of the Binder cumulants are reported in Table 
\ref{tablabinder} and indicate that $\beta_c\simeq0.675$. However, considering
that the curves of the largest lattices have almost the shape of a
``step-function'', and that the intersections take place within the zone
of sudden change of slope, it is difficult to obtain them with good precision.
We have estimated the value of $\beta_c$ by other means, as we shall
see in the next section. 

\begin{table}[tb]
\begin{center}
\begin{tabular}{|c|c|c|c|c|}
\hline
 & 81 & 243 & 729 & 2187\\
\hline
27&0.6693(3)&0.67176(5)&0.67345(3)&0.67455(2)\\
\hline
81&-&0.6730(2)&0.67422(5)&0.67502(1)\\
\hline
243&-&-&0.67486(7)&0.67546(4)\\
\hline
729&-&-&-&0.6759(1)\\
\hline
\end{tabular}
\end{center}
\caption{Values of 
the coupling $\beta$ where the Binder cumulants meet each other.}
\label{tablabinder}
\end{table}

It has been observed \cite{Bonnier} that on fractal lattices,
as the ones we consider,
a certain number of stages of
construction of the SC is necessary in order to obtain
reliable predictions about the
 critical behavior. Actually, as we shall see in the next
sections, we were forced to discard the smallest lattice sizes, $L=27,81$,
in certain fits to obtain reliable results.

\subsection{$\nu$ exponent and $\beta_c(\infty)$}

\subsubsection{Position of the maxima}

In a FSS analysis, every observable, $O$, which diverges in the
thermodynamic limit when $\beta=\beta_c(\infty)$ displays a peak for
finite $L$, located at a different value $\beta_O(L)$. When $L$
increases, the difference between $\beta_O(L)$ and $\beta_c(\infty)$
becomes smaller according to the law
\begin{equation}
\beta_O(L)-\beta_c(\infty)\propto L^{-1/\nu}.
\label{scalingbetacr}
\end{equation}
A three-parameter fit determines the values of exponent $\nu$ and of
$\beta_c(\infty)$.  To produce such a fit we have considered three
different observables: the derivative of the logarithm of the
magnetization, the derivative of the Binder cumulant (\ref{cumulant})
and the isothermal susceptibility (\ref{susc}). These three quantities
display maxima at the values of $\beta_O(L)$ shown in Tables
\ref{maxlogM}, \ref{maxbinder} and \ref{maxsusc}, respectively.

In these Tables (\ref{maxlogM}, \ref{maxbinder} and \ref{maxsusc}) we
report the values of the peaks for the different lattice sizes
together with the statistics performed in correspondence with each $L$.
Here and in the following, each file is made of 200 steps of SW, 400
steps of Metropolis and 2000 cluster sweeps with the Wolff algorithm.
The integrated correlation time is always less than a file 
of measurements.

\begin{table}[tb]
\begin{center}
\begin{tabular}{|r|c|c|c|c|}\hline
$L$ & \# files & $\beta_\mathrm{sim}$ &$\beta_O(L)$ & Max. of 
$(\ln \lan |M| \ran)'(L)$  \\
\hline\hline
27 & $\sim$ 6600 & 0.580 & 0.5796(3) & 13.19(1) \\
\hline
81 & $\sim$ 5800 & 0.629 & 0.6275(1) & 28.46(6) \\
\hline
243 & $\sim$ 4000 & 0.650 & 0.6505(2) & 58.5(2) \\
\hline
729 & $\sim$ 13900 & 0.6618 & 0.6622(1) & 115.4(4) \\
\hline
2187 &  $\sim$ 6400 & 0.6682 & 0.6681(2) & 218(1) \\
\hline
\end{tabular}
\end{center}
\caption{Maxima of $(\ln \lan |M| \ran)'(L)$.}
\label{maxlogM}
\end{table} 

\begin{table}[tb]
\begin{center}
\begin{tabular}{|r|c|c|c|c|}\hline
$L$ & \# files & $\beta_\mathrm{sim}$ &$\beta_O(L)$ & Max. of $U'(L)$ \\
\hline\hline
27 & $\sim$ 6600 & 0.580 & 0.5744(3) & 8.40(2) \\
\hline
81 & $\sim$ 5800 & 0.629 & 0.6252(2) & 17.56(9) \\
\hline
243 & $\sim$ 4000 & 0.650 & 0.6493(2) & 35.8(2) \\
\hline
729 & $\sim$ 13900 & 0.6618 & 0.6617(1) & 69.8(4) \\
\hline
2187 &$\sim$ 2400 & 0.668 & 0.6680(3) & 132(3) \\
\hline
\end{tabular}
\end{center}
\caption{Maxima of $ U'(L)$.}
\label{maxbinder}
\end{table} 

\begin{table}[tb]
\begin{center}
\begin{tabular}{|r|c|c|c|c|}\hline
$L$ & \# files & $\beta_\mathrm{sim}$ &$\beta_{\chi}(L)$ & Max. of 
$\chi(L)$ \\
\hline\hline
27 & $\sim$ 6600 & 0.580 & 0.59431(6) & 27.97(3) \\
\hline
81 & $\sim$ 5800 & 0.629 & 0.63442(4) & 179.2(2) \\
\hline
243 & $\sim$ 2000 & 0.651 & 0.65394(7) & 1164(4) \\
\hline
729 & $\sim$ 8000 & 0.663 & 0.66402(2) & 7759(7) \\
\hline
2187 &$\sim$ 4400 & 0.669 & 0.66936(1) & 51883(58) \\
\hline
\end{tabular}
\end{center}
\caption{Maxima of $\chi(L)$.}
\label{maxsusc}
\end{table} 

The best fits are obtained discarding the $L=27$ data from the three 
Tables. We obtain the results shown in Table \ref{monitor}. All the fits
yield $\chi^2/\dof<1$ (where DF means the number of degrees of freedom
in the fit); we conclude from these that:
\begin{equation}
\beta_c(\infty)=0.6751(1)\quad\quad \nu^{-1}=0.59(1),
\label{beta_nu}
\end{equation}
and
\begin{equation}
\nu=1.70(1) .
\end{equation}

\begin{table}[tb]
\begin{center}
\begin{tabular}{|c|c|c|}\hline
$O$ & $\beta_c(\infty)$ & $\nu^{-1}$ \\
\hline\hline
$\chi$ & 0.67522(7) & 0.589(3) \\
\hline
$(\ln \lan |M| \ran)'$ & 0.6742(5) & 0.62(1) \\
\hline
$U'$ & 0.6747(7) & 0.61(2) \\
\hline
\end{tabular}
\end{center}
\caption{Fits to eq. (\ref{scalingbetacr}) for different observables 
$O$, lattices $81, 243, 729 $ and $2187$.}
\label{monitor}
\end{table}

We remark that we obtain a better result using the susceptibility than
the other two observables (the derivatives of the logarithm of
the absolute value of the magnetization and of the Binder cumulant).

\subsubsection{Fits at fixed values of the Binder cumulant}

Let us now check the values of $\beta_c(\infty)$ and $\nu$ obtained (see
Table \ref{monitor}) employing eq. (\ref{scalingbetacr}) with different
definitions of the apparent critical coupling $\beta_O(L)$.  We consider
now a fixed value $g_0$ of the Binder cumulant $U_L(\beta)$, and define
$\beta(L)$ via the equation
\begin{equation}
U_L(\beta(L))=g_0.
\label{condicion}
\end{equation}
It is clear from reference \cite{PARU5} that $\beta(L)$ behaves like
$\beta_O(L)$, see equation (\ref{scalingbetacr}).

\begin{figure}[htb]
\centerline{\epsfig{figure=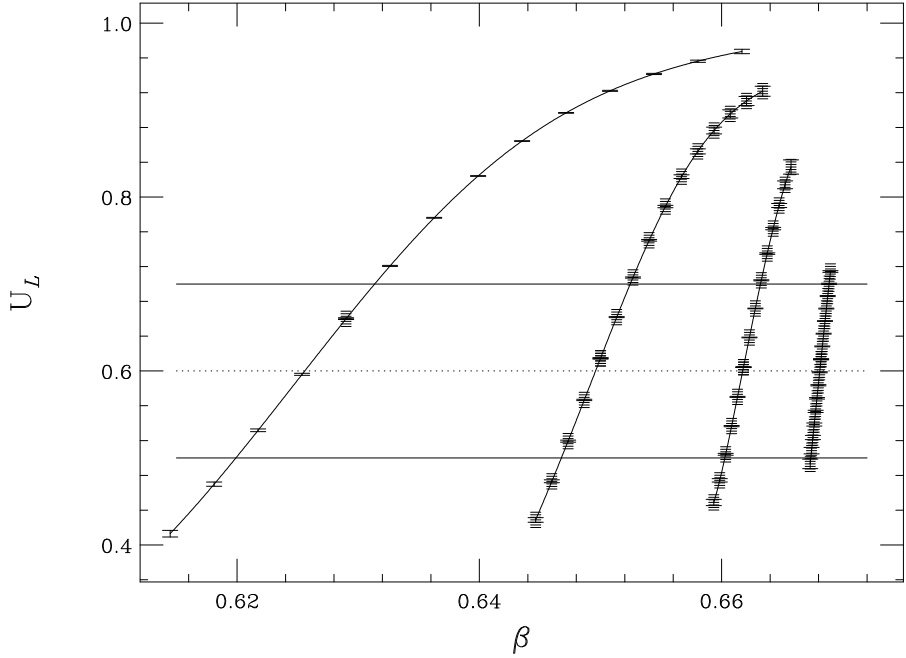,angle=90,width=120mm}}
\caption{Binder cumulants for $L=81,243,729,2187$.}
\label{FIG:BINDER}
\end{figure}

We have used for $g_0$ the values 0.5, 0.6 and 0.7 (see
Fig. \ref{FIG:BINDER}).  The values of the $\beta(L)$ given by
condition (\ref{condicion}) are collected in Table \ref{binders} for
the lattice sizes $L=81,243,729,2187$, and the results of fitting
these values to (\ref{scalingbetacr}) are shown in Table
\ref{fitbinders}. We see that these values of $\beta_c(\infty)$ and
$\nu^{-1}$ are compatible with those given in (\ref{beta_nu}). We
remark that the fits reported in Table \ref{fitbinders} have been obtained
using only the lattice sizes 243, 729 and 2187. In all these fits the number
of degrees of freedom is zero, and the reader should take the results
of the Table \ref{fitbinders} as a check of our previous estimates of
$\beta_c$ and $1/\nu$.

\begin{table}[tb]
\begin{center}
\begin{tabular}{|c||c|c|c|c|}\hline
$g_0$ & $L=81$ & $L=243$ & $L=729$ & $L=2187$ \\
\hline\hline
0.5&0.61972(7)&0.64677(4)&0.66026(2)&0.66736(3)\\ \hline
0.6&0.62551(4)&0.64959(2)&0.661737(6)&0.668099(4)\\ \hline
0.7&0.63134(2)&0.65245(2)&0.663205(9)&0.668868(6)\\
\hline
\end{tabular}
\end{center}
\caption{Values of $\beta(L)$ where the Binder cumulants take
the values 0.5, 0.6 and 0.7.}
\label{binders}
\end{table}

\begin{table}[tb]
\begin{center}
\begin{tabular}{|c|c|c|}\hline
$g_0$ & $\beta_c(\infty)$ & $\nu^{-1}$ \\
\hline\hline
0.5 & 0.6753(2)  & 0.585(6) \\
0.6 & 0.6751(4)  & 0.589(2) \\
0.7 & 0.6752(1)  & 0.584(3) \\
\hline
\end{tabular}
\end{center}
\caption{Fit of the values of Table \ref{binders} to equation 
(\ref{scalingbetacr}).}
\label{fitbinders}
\end{table}

\subsubsection{Peaks of $(\ln \lan |M| \ran)'(L)$ and $U_L'$}

According to the finite size scaling ansatz the derivatives of $(\ln
\lan |M| \ran)(L)$ and $U_L$ must show a peak increasing as
$L^{1/\nu}$.  The values of the maxima in correspondence with each $L$
are reported in Tables \ref{maxlogM} and \ref{maxbinder}.

In the case of $U'_L$ a two-parameter fit gives
\begin{equation}
L>81: \quad  \quad \nu^{-1}=0.600(4) , \quad \chi^2/\dof=1.37,
\end{equation}
which again agrees with our previous result (\ref{beta_nu}).

For $(\ln \lan |M| \ran)'(L)$ we obtain
\begin{equation}
L=729, 2187: \quad \quad \nu^{-1}=0.579(5).
\end{equation}
For $L>27$ or $L>81$ the fits are  bad (i.e. $\chi^2/\dof$ is large).

In these two cases (derivatives of $U$ and $\ln \lan |M| \ran$) 
we have not the necessary accuracy in order to have a stable
fit taking into account the scaling corrections.

\subsection{$\gamma/\nu$ exponent: scaling of $\chi$}
The isothermal susceptibility (\ref{susc}) displays a peak increasing
as $L^{\gamma/\nu}$. 

By fitting the peaks with $L^{\gamma/\nu}$ we obtain a poor $\chi^2/\dof$;
however, after discarding the data relative to $L=27,81$ we obtain a 
$\chi^2/\dof<1$.  The result is:
\begin{equation}
L>81: \quad \gamma/\nu=1.729(1) \quad \chi^2/\dof=0.6 .
\label{fit1gamma}
\end{equation}

In order to include also the data relative to $L=27,81$, we considered
the scaling corrections to such a fit according to
eq. (\ref{omega}). Applying the procedure of the fit
(eq. (\ref{omega})) to the data of Table \ref{maxsusc} (i.e. we
compute the effective exponent $\gamma/\nu$ using all the lattice
sizes: i.e. $\gamma/\nu(27,81)$, $\gamma/\nu(81,217)$,...)
we find:
\begin{equation}
\gamma/\nu=1.76(2)\quad \,  \omega=0.27(14) \quad \ \chi^2/\dof=7,
\end{equation}
which is unsatisfactory. However, if we discard 
only the $L=27$ data we end up with
\begin{equation}
L>27: \quad \gamma/\nu=1.730(1) \quad\  \omega=1.9(8).
\label{fit2gamma}
\end{equation}

\begin{figure}[htb]
\centerline{\epsfig{figure=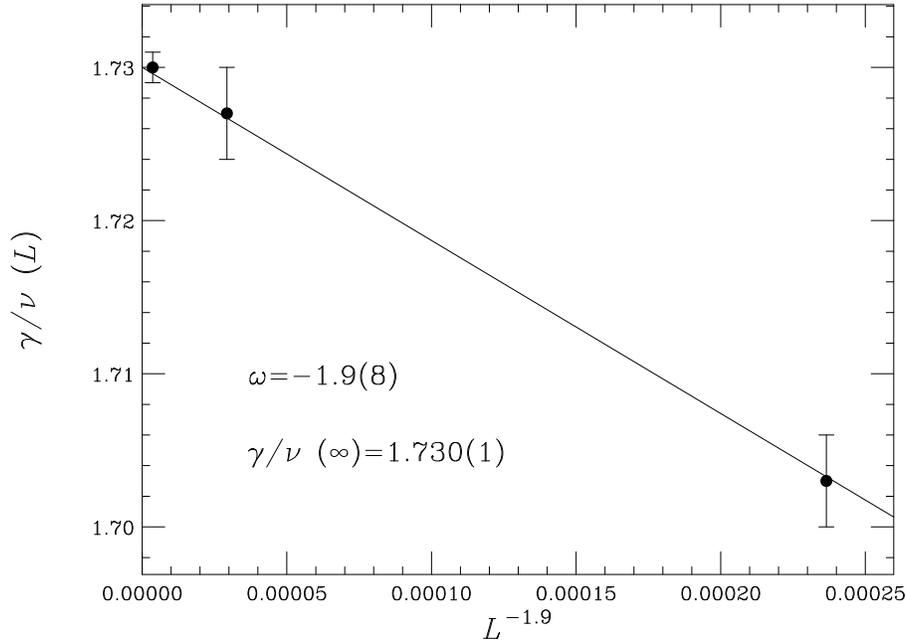,angle=90,width=120mm}}
\caption{Fit to obtain the scaling corrections in $\chi$ using $L=81,243,729,
2187$.}
\label{FIG:FIT}
\end{figure}

In the last fit, shown in Fig. \ref{FIG:FIT}, we have not degrees of freedom,
because we have fitted
three data to (\ref{omega}), which contains three free parameters.
However, the result for $\gamma/\nu$ is compatible with
(\ref{fit1gamma}) having included the $L=81$ data and the exponent
$\omega$ represents the correction to the scaling. It is clear that
the final value of $\gamma/\nu$ in the fit is really stable
(almost-)independently of the $\omega$ value.

We  conclude that:
\begin{equation}
\omega=1.9(8) \quad\quad \, \gamma/\nu=1.730(1).
\label{omegaestim}
\end{equation}
By using $\nu=1.70(1)$ we can write down the result for the $\gamma$
exponent
\be
\gamma= 2.94(2) \ .
\ee

\subsection{$\beta/\nu$ exponent: scaling of $\lan |M| \ran$}

Having computed the critical value of the
coupling (\ref{beta_nu}), we can extract the $\beta/\nu$ exponent from
the scaling law :
\begin{equation}
\lan |M_L| \ran (\beta_c(\infty))\propto L^{-\beta/\nu}.
\label{scalingM}
\end{equation}

We have written in Table \ref{valsM} the values of $\lan |M_L| \ran$ in 
correspondence with the points
$0.6751$, $0.6750$ and $0.6752$, and the final value
 that we take to make the fit (that is, it takes into account the
error in our final value of $\beta_c$).

\begin{table}[tb]
\begin{center}
\begin{tabular}{|r|c|c|c||c|}\hline
$L$ & $\lan |M_L| \ran(0.6751)$ & $\lan |M_L| \ran (0.6750)$ 
& $\lan |M_L| \ran (0.6752)$ & $\lan |M_L| \ran(\beta_c(\infty))$\\
\hline\hline
27 & 0.8546(2) & 0.8544(2) & 0.8548(2) & 0.8546(4) \\
\hline
81 & 0.7945(1) & 0.7942(1) & 0.7949(1) & 0.7945(5) \\
\hline
243 & 0.7345(1) & 0.7338(1) & 0.7351(1) & 0.7345(8) \\
\hline
729 & 0.6742(1) & 0.6729(1) & 0.6754(1) & 0.6742(14) \\
\hline
2187 & 0.61338(8) & 0.61110(8) & 0.61561(9) & 0.61336(234) \\
\hline
\end{tabular}
\end{center}
\caption{Absolute value of the magnetization at the critical point 
for the different lattice sizes.}
\label{valsM}
\end{table} 

Only by fitting the lattice sizes with $L\ge 243$ it is possible to
obtain a ``reasonable'' fit ($\chi^2/\dof=2.4/1$; we have only one
degree of freedom), in particular
\be
\beta/\nu=0.080(1).
\ee

The quality of the data is not good for trying a fit taking into
account the scaling corrections.

\subsection{$d_{\mathrm{s}}$ and hyperscaling}

Having obtained the values of $\gamma/\nu=1.730(1)$
and $\beta/\nu=0.080(1)$ we can consider the hyperscaling relation
\begin{equation}
2 \frac{\beta}{\nu}+\frac{\gamma}{\nu}=d,
\label{hyprel}
\end{equation}
where $d$ is the dimension that controls hyperscaling. In our case,
we obtain $d=1.890(2)$, which is close  to the Hausdorff dimension
of the lattice under scrutiny.

Finally we remark that it is clear from the standard derivation of the
hyperscaling relations that the dimension that plays a role is the 
Hausdorff one.

\subsection{$\alpha/\nu$ exponent: the specific heat}

The specific heat, for positive values of $\alpha$, has a peak which
scales with $L^{\alpha/\nu}$.  In our case, this peak is very flat
showing a plateau that goes down as the lattice size increases. This
corresponds to a negative value of the exponent $\alpha$. Our numerical
estimate is $\alpha\simeq -1.15$ (which derives from $\nu d_H=2
-\alpha$). Notice that such a value lies between the zero value of
the two dimensional Ising model and the infinitely negative value of
the one dimensional Ising model.

As expected the exponents for the present lattice
interpolate between the corresponding values of the $d=1$ and $d=2$ cases
as shown 
in Table \ref{tablaexact}. The value of the $\omega$ exponent in $d=1$ is
calculated in the Appendix.

\begin{table}[tb]
\begin{center}
\begin{tabular}{|c|c|c|c|c|}
\hline
 & $\beta_c$ & $\nu$ & $\gamma$ & $\omega$ \\
\hline\hline
$d=1$ & $\infty$ & $\infty$ & $\infty$ & 2 \\
\hline
this fractal & 0.6752(1) & 1.70(1) & 2.94 (2) & 1.9(8) \\ 
\hline
$d=2$ & 0.44... & 1 & 1.75 & 4/3 \\ 
\hline
\end{tabular}
\end{center}
\caption{Comparison between the results for the phase transition in the
$d=1$, $d=2$ and $d=1.89$ (fractal A) Ising models.}
\label{tablaexact}
\end{table}

\section{Numerical results for the fractal B}

In order to provide further evidence for the hyperscaling
relation we considered a second fractal lattice, the Sierpinski carpet B,
described in section 2, whose Hausdorff dimension
is $d_H=\ln 12/\ln 4$. We carried out a finite size scaling analysis
with lattice sizes of $L=64,256,1024$ and $4096$. We were interested in
the verification of the hyperscaling relation (\ref{hyprel}). In order to
get an estimate of the critical exponents $\gamma/\nu$ and $\beta/\nu$ 
from a single simulation for every size $L$, we considered in this case 
the observables
$\chi$ and the derivative of the magnetization with respect to $\beta$,
as it is explained in the following paragraphs. The results of the
simulations are shown in Table~\ref{secondfr}.

\begin{table}[tb]
\begin{center}
\begin{tabular}{|r|c|c|c|c|}\hline
$L$ & $\beta_\mathrm{sim}$ &$\beta_c(L)$ & 
Max. of $d M /d \beta$ & Max. of $\chi(L)$ \\
\hline\hline
64 & 0.7396 & 0.74548(6) & 5.797(3) & 86.57(2) \\
\hline
256 & 0.8060 & 0.8094(3) & 9.06(2) & 839.8(5) \\
\hline
1024 & 0.8480 & 0.8501(6) & 12.67(5) & 8484(3) \\
\hline
4096 & 0.8780 & 0.8782(6) & 16.6(6) & 88808(101) \\
\hline
\end{tabular}
\end{center}
\caption{Results of the simulation of the  fractal B.}
\label{secondfr}
\end{table} 

The exponent $\gamma/\nu$ is obtained from the peak of the susceptibility. 
A fit with the
three first lattice sizes gives the value $\gamma/\nu=1.6530(2)$, and
the value $\gamma/\nu=1.6755(4)$ with the three last ones. This difference
means the existence of scaling corrections. 
We shall take as our best estimate the value
\begin{equation}
\gamma/\nu=1.67(2).
\label{gnuexp}
\end{equation}

Differentiating the magnetization with respect to $\beta$, we obtain a
quantity which scales
as $L^{(1-\beta)/\nu}$. Discarding again the $L=64$ data, we find:
\begin{equation}
\frac{1-\beta}{\nu}=0.241(3) \quad\quad \chi^2/\mathrm{DF}=3.1.
\label{expx}
\end{equation}
The large value of $\chi^2/\mathrm{DF}$ again shows the existence of
large scaling corrections. However, using only the data of the lattice
sizes $L=1024$ and $L=4096$ we have found $(1-\beta)/\nu=0.20(3)$
that is compatible in the error bars with (\ref{expx}).

\begin{figure}[htb]
\centerline{\epsfig{figure=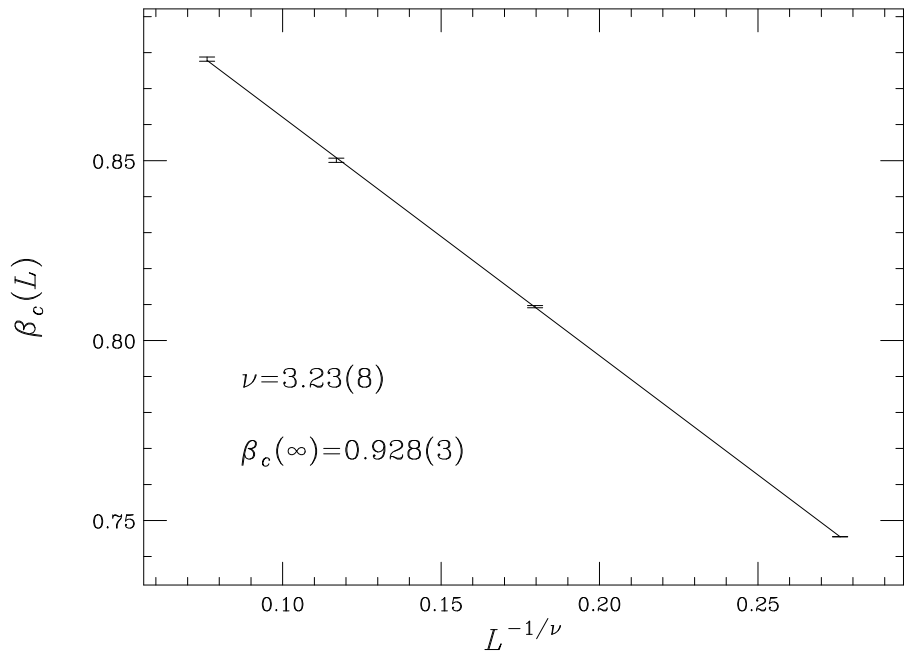,angle=90,width=120mm}}
\caption{Best fit to $\beta_c(\infty)$ obtained when varying $\nu$, 
using the values of $\beta_c(L)$ shown in Table~\ref{secondfr}. The error
in $\beta_c(\infty)$ contains the uncertainty in $\nu$.}
\label{FIG:FIT2}
\end{figure}

Now we need an estimate of the $\nu$ exponent in order to get $\beta/\nu$
from (\ref{expx}). We have obtained a good fit to eq. (\ref{scalingbetacr})
using the positions of the peaks of the derivative of $M$ with respect to
$\beta$. These are shown as $\beta_c(L)$ in Table~\ref{secondfr}. In this
case we are dealing with a three-parameter fit, so that we need to use the
four lattice sizes for consistency. Letting $\nu$ vary freely, we obtain 
the best fit for
\begin{equation}
\nu=3.23(8) \quad \beta_c(\infty)=0.928(3) \quad \chi^2/\mathrm{DF}=0.95.
\end{equation}
This fit is shown in Fig.~\ref{FIG:FIT2}.
Using this result for $\nu$ and (\ref{expx}), we get
\begin{equation}
\beta/\nu=0.069(10).
\label{bnuexp}
\end{equation}

With the results (\ref{gnuexp}) and (\ref{bnuexp}), we obtain through
the hyperscaling relation (\ref{hyprel}), the value $d=1.81(3)$. The
error in this value is large owing to the scaling corrections, but it
is perfectly compatible with the Hausdorff dimension of the fractal.

\section{Conclusions}
In the present paper we have considered the issue of phase transitions
on discrete fractal lattices. 
First, by means of an exact Migdal-Kadanoff
decimation transformation we confirmed the absence of
a finite temperature critical point for the Sierpinski gasket.

Then we focused on lattices with infinite ramification order and
performed  extensive MC simulations on deterministic 
Sierpinski  carpets of different sizes and different
Hausdorff dimensions, where we observed the presence of
a finite temperature phase transition.
 The FSS analysis was applied in order to
extract from the data the critical exponents and their numerical errors.  

 To the best of our knowledge the present study represents the
first massive attempt to determine numerically the critical properties of 
Ising models on fractal 
lattices. We have illustrated the difficulties one encounters when
dealing with these systems, which are due to the slow 
convergence of the averages to the infinite volume limit.

Interestingly, the hyperscaling relation, which was not obtained before
for such lattices,
indicates that the role of $d$ of the Euclidean case is taken by the
Hausdorff dimension.
Finally we should comment about the universality of the present
results. As pointed out by Wu and Hu \cite{Hu}, on a true self similar
structure the statement of universality survives only in a weak version.
In fact, the various exponents might depend on the detailed structure of 
the fractal. It has been found, by means of approximate treatments,
that the exponents vary even when the fractal dimension remains fixed,
because they depend also on other geometrical characteristics
such as the connectivity and the lacunarity. This is not too surprising
since due to the scale invariance a small change at the smallest stage
may be magnified up to the largest one. In other words, the lattice structure,
which in normal critical phenomena does not enter with all its details
here plays a much more important role.
However, we have verified that the hyperscaling relation holds for the
two fractals we have considered, which suggests that for lattices having
the same fractal dimension, $\beta$, $\nu$ and $\gamma$ may vary with the
lacunarity, but the sum of their ratios cannot!

We finally remark that the model  under study  could provide  a very
good benchmark to study the thermodynamics of the Ising  model near its
lower critical dimension.

\section*{\protect\label{S_ACKNOWLEDGES}Acknowledgments}

We wish to thank L.A. Fern\'andez for very useful discussions.
J.M. Carmona is a Spanish MEC fellow.
J. J. Ruiz-Lorenzo is supported by an EC HMC (ERBFMBICT950429) grant.
The numerical work was done using the RTNN parallel machine (composed
by 32 Pentium Pro processors) located at Zaragoza University.

\section{Appendix}

In this appendix we will compute the correction-to-scaling
exponent for the one dimensional Ising model.

We will use the Binder cumulant (\ref{cumulant}), which can be written as
\be
U_L(\beta)=-\frac{\lan M^4\ran_c}{2\lan M^2\ran_c^2}\ ,
\label{cum}
\ee
where 
\be
\lan M^n\ran_c=\left.\frac{\partial^n\ln{\cal Z}_V}
{\partial h^n}\right|_{h=0},
\label{sconexo}
\ee
with $h$ the magnetic field and ${\cal Z}_V$ the partition function
for the volume $V$ (in this case, $V\equiv L$).

It is possible to diagonalize exactly the transfer matrix for the one
dimensional model, and the two eigenvalues are
\be
\lambda_{1,2}= e^\beta \left[\cosh(h) \pm \left(\sinh^2(h) +e^{-4
\beta} \right)^{1/2}\right].
\ee
Then the  partition function is
\be
{\cal Z}_V=\lambda_1^V +\lambda_2^V \ ,
\label{particion}
\ee

>From (\ref{sconexo}) and (\ref{particion}), we obtain 
\be
\lan M^2\ran_c=V\,e^{2\beta}\,\frac{1-\left(\tanh(\beta)\right)^V}
{1+\left(\tanh(\beta)\right)^V}\,,
\ee
and
\be
\lan M^4\ran_c=\frac{V e^{2\beta}\left(3 e^{4\beta}-1\right)
        \left(\left(2\sinh(\beta)\right)^{2V}-
              \left(2\cosh(\beta)\right)^{2V}\right)
        +12 V^2 e^{4\beta}
        \left(4\sinh(\beta)\cosh(\beta)\right)^V}
        {\left(\left(2\sinh(\beta)\right)^V+
        \left(2\cosh(\beta)\right)^V\right)^2}\ .
\ee     

\begin{figure}[tb]
\centerline{\epsfig{figure=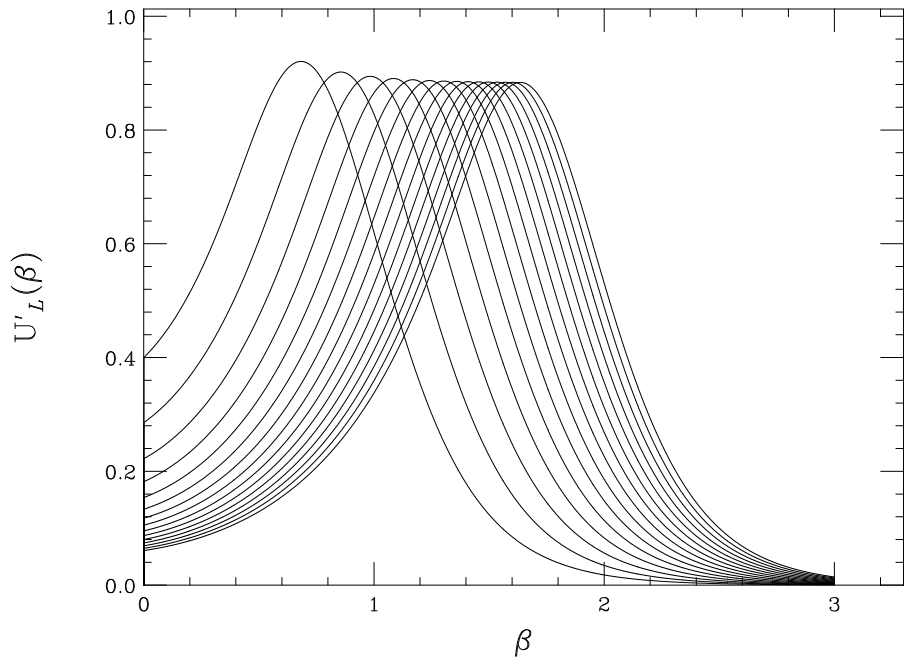,angle=90,width=120mm}}
\caption{Derivative of the Binder cumulant as a function of $\beta$
for several values of $L=4,8,16,32,64,128...$ for the Ising model
in one dimension.}
\label{FIG:OMEGA1d}
\end{figure}

We put these expressions into (\ref{cum}), and consider the derivative of
the Binder cumulant. It has a peak at finite values of $\beta$ for each $L$,
$\beta_c(L)$.
The peak of the derivative of the Binder cumulant scales as $L^{1/\nu}$
plus scaling corrections. In $d=1$, $\nu=\infty$, so that we have
\be
U'_L(\beta_c(L))=A+B L^{-\omega}+ O(L^{-2\omega}).
\label{corrsc}
\ee
We calculated numerically this derivative and obtained the plot shown
in Fig. \ref{FIG:OMEGA1d} for several values of $L$. 

Fitting to equation (\ref{corrsc}) we obtain, when $L\to\infty$, the
asymptotic limit
\be
\omega=2.
\ee


\begin{thebibliography}{99}

\bibitem{note}
Deterministic fractal lattices and phase transitions have been
studied in the past \cite{gefen}, \cite{spheric,Io} by different
methods. In particular \cite{gefen} used an approximate
Migdal-Kadanoff bond moving renormalization group approach
and obtained results for the critical temperature
of both the Sierpinski gasket and the Sierpinski carpet.

\bibitem{spheric}
The spherical model has been recently studied by
U. Marini Bettolo Marconi and A. Petri
Phys. Rev. E  {\bf 55}, 1311 (1997); J.Phys. A   {\bf 30}, 1069 (1997);
Phil. Mag. {\bf 77B}, 265 (1998).

\bibitem{Io}
U. Marini Bettolo Marconi.
Physical Review E {\bf 57}, 1290 (1998).

\bibitem{PARU} 
G. Parisi and J. J. Ruiz-Lorenzo, 
J.  Phys. A: Math and Gen {\bf 28}, L395 (1995). 

\bibitem{DILU4} 
H. G. Ballesteros et al., hep-lat/9707017, Nucl. Phys. B (in press).

\bibitem{DILU2} 
J. J. Ruiz-Lorenzo, J. Phys. A: Math and Gen  {\bf 30}, 485 (1997);
 H. G. Ballesteros et al., J. Phys. A: Math and Gen {\bf 30}, 8379 (1997). 

\bibitem{STAUFFER}
D. Stauffer  and A. Aharony, {\em Introduction to the percolation
theory}, Taylor and Francis 1994 (2nd revised edition).

\bibitem{voids}
The lacunarity serves to measure the failure of a fractal to be
translationally invariant. It has been found that in the Migdal-Kadanoff
renormalization scheme the values of the exponents vary not only
with the fractal dimension, but also with the connectivity and lacunarity.
According to reference \cite{gefen}, $d_H$ and ${\cal{R}}$ fixed, the exponent
$\nu$ increases with decreasing lacunarity.

\bibitem{PaMa} G. Martinelli and G. Parisi, Nucl. Phys. B {\bf 180 (FS)}
201 (1981).

\bibitem{formula} We recall that this formula has been obtained for integer
dimensions and we use its analytical prolongation to non integer
dimensions. Do not confuse with the Migdal-Kadanoff techniques used in
reference \cite{gefen}, in which the authors worked directly on the fractals.

\bibitem{Brezin} E. Br\'ezin, J. Physique {\bf 43}, 15 (1982).

\bibitem{Barber} M.E. Fisher and M.N. Barber, Phys. Rev. Lett. {\bf 28}, 
1516 (1972).

\bibitem{SDM} M. Falcioni, E. Marinari, M. L. Paciello, G. Parisi and
B. Taglienti, Phys. Lett B {\bf 108}, 331 (1982); A. M. Ferrenberg and
R. H. Swendsen, Phys. Rev. Lett. {\bf 61}, 2635 (1988).

\bibitem{WOLFF} U. Wolff, Phys. Rev. Lett. {\bf 62}, 3834 (1989).

\bibitem{SW} R. H. Swendsen and J. S. Wang, Phys. Rev. Lett. {\bf 58}, 
86 (1987).

\bibitem{Bonnier} B. Bonnier, Y. Leroyer and C. Meyers, 
Phys. Rev B {\bf 37}, 5205 (1988).

\bibitem{PARU5} G. Parisi and J. J. Ruiz-Lorenzo, 
Phys. Rev. B {\bf 54} R3698 (1996). Erratum: ibid {\bf 55}, 6082 (1997).

\bibitem{Hu} Y. Wu and B. Hu, Phys.Rev. A {\bf 35}, 1404 (1987).

\bibitem{gefen}
Y. Gefen, A. Aharony and B. B. Mandelbrot, J. Phys. A {\bf 17}, 1277 (1984)
and references therein.


\end{thebibliography}
\end{document}